\documentclass[11pt]{article}
\usepackage{amsmath,amssymb,color,cite}
\usepackage{graphicx}
\usepackage{amsfonts}
\usepackage{enumerate}  
\usepackage{bbm}
\usepackage{multirow}
\usepackage{nicefrac}
\usepackage[all]{xy}
\usepackage{booktabs}
\usepackage{amsfonts}
\usepackage{comment}
\usepackage{slashed}
\usepackage[section]{placeins}

\def\tsz{.9}

\usepackage{float} 

\def\ad{{\dot \alpha}}

\def\a{\alpha}
\def\b{\beta}
\def\c{\gamma}

\def\d{\delta}

\def\e{\epsilon}
\def\ve{\varepsilon}

\def\x{\xi}

\def\3h{\frac32}
\def\1h{\frac12}
\def\lb{\left(}
\def\rb{\right)}

\def\cO{{\cal O}}

\def\nn{\nonumber}
\def\ed{\end{document}}
\def\ba{\begin{array}}
\def\ea{\end{array}}
\def\bea{\begin{eqnarray}}
\def\eea{\end{eqnarray}}
\def\ft#1#2{{\textstyle{{\scriptstyle #1}
\over {\scriptstyle #2}}}}

\newcommand{\be}{\begin{equation}}
\newcommand{\ee}{\end{equation}}
\newcommand{\eq}[1]{(\ref{#1})}

\newcommand{\w}[1]{\\[0.#1cm]}

\def\det{{\rm det\,}}

\def\ob{{\overline\Box}}
\def\on{{\overline\nabla}}

\def\ta{\small
\begin{table}[ht]
\begin{center}
\setlength{\tabcolsep}{10pt} 
\renewcommand{\arraystretch}{1.5} 
\begin{tabular}{lcll}
\hline
Fields  &   $SO(4)\times SO(7)$ content  & $SO(5)\times SO(8)$ content & Restrictions  \\
\hline
$H_{\mu\nu}$ &  $(20)(000)$   & $(n2)(\ell)$\ ,\ $(n)(\ell)$ \ ,\  $(n1)(\ell)$ & $n\ge 2\ , \ell\ge 0$ 
\\
$M$ &    $(00)(000)$   & $(n)(\ell)$ & $n\ge 0\ , \ell\ge 0$ 
\\
$K_{\mu\a}$ &    $(10)(100)$   & \  $(n1)(\ell 1)$\ ,\ $(n)(\ell)$ ,\ $(n1)(\ell)$\ , $(n)(\ell 1)$  & $n\ge 1\, \ell\ge 1$ 
\\
$L_{\a\b} $ &   $(00)(200)$   & $(n)(\ell 2)$ \ , \ $(n)(\ell)$\ ,\  $(n)(\ell 1)$  &  $n\ge 0\ , \ell\ge 2$
\\
$N$ & $(00)(000)$   & $(n)(\ell)$ &  $n\ge 0\ , \ell\ge 0$ \\
$W_\mu$ & $(10)(000)$   &  $(n)(\ell)$ \ ,\ $(n1)(\ell)$  & \\
$X_{\mu\nu\a}^\pm $ & $(1, \pm1)(100)$   & $(n1)(\ell 1)$\ ,\ $(n1)(\ell)$ & $n\ge 1\ , \ell\ge 0$\\
$Y_{\mu\a\b}$ & $(10)(110)$   & $(n1)(\ell 11)$\ , \ $(n 1)(\ell 1)$\ ,\ $(n)(\ell 1)$\ ,\ $(n)(\ell 1 1)$ & $n\ge 1\ , \ell\ge 1$ \\
$Z_{\a\b\c}$ &  $(00)(111)$   & $(n)(\ell 11, \pm 1)$\ ,\ $(n)(\ell 11)$ & $n\ge 0\ , \ell \ge 1$ \\
$\eta_{r\a}$ & $\left(\3h,\1h\right)\left(\1h,\1h,\1h\right)$ & $\left(n,\3h\right)\left(\ell,\1h,\1h,\pm \1h\right)$\ ,\ $\left(n,\1h\right)\left(\ell,\1h,\1h\, \pm\1h\right)$ & $n\ge \3h\ , \ell\ge \1h$ 
\\
$\eta_{r\ad}$ & $\left(\3h,-\1h\right)\left(\1h,\1h,\1h\right)$ & $\left(n,\3h\right)\left(\ell,\1h,\1h,\pm \1h\right)$\ ,\ $\left(n,\1h\right)\left(\ell,\1h,\1h\,\pm\1h\right)$ & $n\ge \3h\ , \ell\ge \1h$ 
\\
$\chi_{i\a}$ & $\left(\1h,\1h\right)\left(\3h,\1h,\1h\right)$ & $\left(n,\1h\right)\left(\ell,\3h,\1h,\pm \1h\right)$\ ,\ $\left(n,\1h\right)\left(\ell,\1h,\1h,\pm\1h\right)$ & $n\ge \1h\ , \ell\ge \3h$ 
\\
$\chi_{i\ad}$ & $\left(\1h,-\1h\right)\left(\3h,\1h,\1h\right)$ & $\left(n,\1h\right)\left(\ell,\3h,\1h,\pm \1h\right)$\ ,\ $\left(n,\1h\right)\left(\ell,\1h,\1h,\pm\1h\right)$ & $n\ge \1h\ , \ell\ge \3h$  
\\
$\lambda_{\a}$ &  $\left(\1h,\1h\right)\left(\1h,\1h,\1h\right)$ & $\left(n,\1h\right)\left(\ell,\1h,\1h,\pm \1h\right)$ & $n\ge \1h\ , \ell\ge \1h$
\\
$\lambda_{\ad}$ & $\left(\1h,-\1h\right)\left(\1h,\1h,\1h\right)$ & $\left(n,\1h\right)\left(\ell,\1h,\1h,\pm \1h\right)$ & $n\ge \1h\ , \ell\ge \1h$
\\
\hline
\end{tabular}
\end{center}
\caption{{\small The highest weights of the $G$ representations occurring in the harmonic expansions of the fields listed in the first columns which have the $H$-representation content listed in the second column. In representations $(nn_1)(\ell\ell_1 \ell_2 \ell_3)$ it is understood that $n\ge n_1$ and $\ell \ge \ell_1\ge \ell_2 \ge |\ell_3|$. 
The last entries in each row, and the second to the last entries for $K_{\mu\a}$ and $Y_{\mu\a}$ turn out to be unphysical gauge modes. There are six $(n)(\ell)$ entries but only four of them are independent. We have chosen to eliminate those in $H_{\mu\nu}$ and $L_{\a\b}$.
 }}
\label{GH}
\end{table}}

\def\fa{
\begin{figure}[ht]
\begin{center}
\unitlength=\tsz mm
\framebox{
\begin{picture}(170,130)(-19,-10)
\put(0,100){\makebox(0,0){  \framebox{$\ve_0 :=\frac{\ell}{2}+3$}}}
\put(70,100){\makebox(0,0){$(\ve_0,2)(\ell 000)$}}
\put(45,100){\makebox(0,0){$H_{\mu\nu}\,:$}}
\put(70,95){\vector(0,1){0}}
\put(70,95){\vector(0,-1){10}}
\put(70,80){\makebox(0,0){$ \lb \ve_0 \mp \1h,\frac32 \rb \lb \ell\pm\frac12,\frac12,\frac12,\pm \frac12 \rb $}}
\put(35,80){\makebox(0,0){$\eta_\mu\,:$}}
\put(60,75){\vector(-1,-2){5}}
\put(60,75){\vector(1,2){0}}
\put(80,75){\vector(1,-2){5}}
\put(80,75){\vector(-1,2){0}}
\put(50,60){\makebox(0,0){$(\ve_0,1)(\ell 1 1 0)$}}
\put(25,60){\makebox(0,0){$Y_{\mu\a\b}\,:$}}
\put(50,55){\vector(-1,-2){5}}
\put(50,55){\vector(1,2){0}}
\put(60,55){\vector(2,-1){20}}
\put(60,55){\vector(-2,1){0}}
\put(90,60){\makebox(0,0){$(\ve_0\mp 1,1)(\ell\pm 1,1,0,0)$\,:}}
\put(135,60){\makebox(0,0){$(K_{\mu\a}, X_{\mu\nu\a})$}}
\put(85,55){\vector(-2,-1){20}}
\put(85,55){\vector(2,1){0}}
\put(90,55){\vector(1,-2){5}}
\put(90,55){\vector(-1,2){0}}
\put(40,40){\makebox(0,0){$(\ve_0\mp \3h,\frac12)(\ell\pm \frac32,\frac12,\frac12,\mp \frac12)^\star $}}
\put(-2,40){\makebox(0,0){$(\partial\eta, \partial\chi, \lambda)\,:$}}
\put(40,35){\vector(-1,-2){9}}
\put(40,35){\vector(1,2){0}}
\put(50,35){\vector(2,-3){12}}
\put(50,35){\vector(-2,3){0}}
\put(100,40){\makebox(0,0){($\ve_0 \mp \1h,\frac12)(\ell\pm \frac12, \frac32, \frac12,\mp \frac12)\,:$}}
\put(135,40){\makebox(0,0){$\chi_\a$}}
\put(92,35){\vector(-2,-3){12}}
\put(92,35){\vector(2,3){0}}
\put(100,35){\vector(-1,2){0}}
\put(100,35){\vector(1,-2){9}}

\put(20,12){\makebox(0,0){$(\ve_0 \mp 2,0)(\ell \pm 2,0,0,0)^\star $}}
\put(20,2){\makebox(0,0){$(M,N,\partial W,\partial\partial K)$}}
\put(70,12){\makebox(0,0){$ (\ve_0\mp 1,0)(\ell\pm 1,1,1,\mp 1) $}}
\put(70,2){\makebox(0,0){$ Z_{\a\b\c} $}}
\put(120,12){\makebox(0,0){$ (\ve_0,0)(\ell,2,0,0) $}}
\put(120,2){\makebox(0,0){$ L_{\a\b} $}}
\end{picture}
}
\end{center}
\caption{\small The spectrum of $11D$ supergravity on $AdS_4 \times S^7$. The $11D$ supergravity fluctuation fields from which the states come from are defined in \eq{sdefs}, and shown in the figure. Here $\mu=0,1,...,3$ labels the $AdS_4$ spacetime, and $\a=1,...,7$ labels the $S^7$  coordinates. The representations are labelled by the highest weights $(E_0,s)(\ell, \ell_1,\ell_2,\ell_3)$, where $E_0 \ge s$ and $\ell\ge \ell_1 \ge \ell_2\ge |\ell_3|)$. The corresponding Dynkin labels for $SO(8)$ are $(\ell-\ell_1,\ \ell_1-\ell_2,\ \ell_2-\ell_3,\ \ell_2+\ell_3)$. Each value of $\ell$ gives a $OSp(8|4)$ multiplet. $\ell=0$ gives the massless $4D$ maximal supergravity multiplet. for which $E_0=s+1$, while  $\ell=-1$ gives the singleton multiplet, contained in the towers marked by $\star$ in the figure. States for $\ell\ge 1$ are massive multiplets with $d_\ell \times (128_B+128_F)$ degrees of freedom, where $d_\ell$ is the dimension of the $\ell$'th rank totally symmetric and traceless $SO(8)$ tensor. One can define $m^2_B$ and $m^2_F$ for bosons and fermions, respectively, such that they actually vanish for the $AdS_4$ massless states with $s=0,1/2,1$, as follows: $m_B^2=4E_0(E_0-3)+8$ and $m_F^2=(2E_0-3)^2$.}
\end{figure}
\label{AdS4}
}

\def\fb{
\begin{figure}[ht]
\begin{center}
\unitlength=\tsz mm
\framebox{
\begin{picture}(170,130)(-19,-10)
\put(4,100){\makebox(0,0){ \framebox{$\e_0 :=2n+6$} }}
\put(35,100){\makebox(0,0){$L_{\a\b}\,:$}}
\put(60,100){\makebox(0,0){$(\e_0,2,0,0)(n,0)$}}
\put(60,95){\vector(0,1){0}}
\put(60,95){\vector(0,-1){10}}
\put(60,80){\makebox(0,0){ $\lb \e_0\mp \1h,\frac32,\frac12,\mp \frac12 \rb \lb n\pm \1h, \frac12\rb $}}
\put(25,80){\makebox(0,0){ $\chi_\a\,:$}}
\put(40,75){\vector(-1,-2){5}}
\put(40,75){\vector(1,2){0}}
\put(60,75){\vector(1,-2){5}}
\put(60,75){\vector(-1,2){0}}
\put(90,75){\vector(4,-1){40}}
\put(90,75){\vector(-4,1){0}}
\put(20,60){\makebox(0,0){$ (\e_0, 1,1,0)(n,1)$}}
\put(-8,60){\makebox(0,0){$Y_{\mu\a\b}\,: $}}
\put(70,60){\makebox(0,0){$(\e_0\mp 1,1,0,0)(n\pm 1,1)$}}
\put(80,70){\makebox(0,0){$(K_{\mu\a}, X_{\mu\nu\a})$}}
\put(125,60){\makebox(0,0){$(\e_0\mp 1,1,1,\mp 1)(n\pm 1,0)^\star $}}
\put(130,70){\makebox(0,0){$Z_{\a\b\c}$}}
\put(30,55){\vector(-1,-2){5}}
\put(30,55){\vector(1,2){0}}
\put(70,55){\vector(1,-2){5}}
\put(70,55){\vector(-1,2){0}}
\put(60,55){\vector(-2,-1){20}}
\put(60,55){\vector(2,1){0}}
\put(40,55){\vector(2,-1){20}}
\put(40,55){\vector(-2,1){0}}
\put(110,55){\vector(-3,-2){10}}
\put(110,55){\vector(3,2){0}}
\put(25,40){\makebox(0,0){$\lb \e_0 \mp \1h,\1h,\1h,\pm \1h\rb\lb n \pm \1h,\3h\rb $}}
\put(-10,40){\makebox(0,0){$\eta_\mu\,:$}}
\put(90,40){\makebox(0,0){$\lb \e_0 \mp \3h,\1h,\1h,\mp \1h\rb\lb n \pm \3h,\frac12\rb^\star
\,: $}}
\put(135,40){\makebox(0,0){$(\partial\eta, \partial\chi, \lambda)$}}
\put(30,55){\vector(-1,-2){5}}
\put(30,55){\vector(1,2){0}}
\put(70,55){\vector(1,-2){5}}
\put(70,55){\vector(-1,2){0}}
\put(80,35){\vector(1,-2){5}}
\put(80,35){\vector(-1,2){0}}
\put(20,35){\vector(-1,-2){5}}
\put(20,35){\vector(1,2){0}}

\put(10,20){\makebox(0,0){$ (\e_0,0,0,0)(n,2) $}}
\put(10,10){\makebox(0,0){$ H_{\mu\nu}$}}
\put(85,20){\makebox(0,0){$ (\e_0\mp 2,0,0,0)(n\pm2,0)^\star $}}
\put(85,10){\makebox(0,0){$ (M,N,\partial W,\partial\partial K)$}}
\put(30,55){\vector(-1,-2){5}}
\put(30,55){\vector(1,2){0}}
\put(70,55){\vector(1,-2){5}}
\put(70,55){\vector(-1,2){0}}
\end{picture}
}
\end{center}
\caption{\small The spectrum of $11D$ supergravity on $AdS_7 \times S^4$. The $11D$ supergravity fluctuation fields from which the states come from are defined in \eq{sdefs}, and shown in the figure. Here $\a=0,1,...,6$ labels the $AdS_7$ spacetime, and $\mu=1,...,4$ labels the $S^4$ coordinates. The representations are labelled by the highest weights of $SO(6,2)\times SO(5)$ as $(E_0, \ell_1,\ell_2,\ell_3)(n,n_1)$, where $E_0\ge \ell_1 \ge \ell_2\ge |\ell_3|$ and $n \ge n_1$. The Dynkin labels for $SO(5)$ are $(n-n_1,2n_1)$ and the corresponding Dynkin labels for $USp(4)$ are $(2n_1, n-n_1)$. Each value of $\ell$ gives an $OSp(6,2|4)$ multiplet. $n=0$ gives the $7D$ maximal supergravity multiplet, for which $E_0=\ell_1+4$, saturating the unitarity bound.  $n=-1$ gives the doubleton multiplet. These sit in the towers marked by $\star$ in the Figure. States for $n \ge 1$ are massive multiplets with $d_n \times (128_B+ 128_F)$ degrees of freedom, where $d_n$ is the dimension of $n$'th rank totally symmetric and traceless $SO(5)$ tensor.}
\label{AdS7}
\end{figure}
}

\begin{document}

\begin{flushright}
\hfill{MI-TH-201}\\    


\end{flushright}

\vspace{25pt}

\begin{center}
{\Large {\bf 11D Supergravity on $AdS_4 \times S^7$ versus  $AdS_7 \times S^4$}}

\vspace{0.3in}

{\large Ergin Sezgin}

\vspace{0.3in}

{\it George and Cynthia Woods Mitchell  Institute
for Fundamental Physics and Astronomy,\\
Texas A\&M University, College Station, TX 77843, USA}

\vspace{40pt}

\begin{abstract}

The maximally supersymmetric Freund-Rubin vacua for eleven dimensional supergravity, namely $AdS_4 \times S^7$  and $AdS_7 \times S^4$, admit an analytic continuation to $S^4 \times S^7$. From the full harmonic expansions on $S^4 \times S^7$, it is shown that by analytical continuation to either $AdS_4$, or to $AdS_7$, the detailed structure of the Kaluza-Klein spectrum can be obtained for both vacua in a unified manner. The results are shown to be related by a simple rule which interchanges the spacetime and internal space representations. We also obtain the linearized field equations for the singletons and doubletons but they can be gauged away by fixing certain Stuckelberg shift symmetries inherited from the Kaluza-Klein reduction.

\end{abstract}

\end{center}

\vspace{15pt}

\thispagestyle{empty}

\pagebreak
\voffset=-40pt

\section{Introduction}

This paper is dedicated to the memory of Peter Freund.  I did not have the opportunity to work with him but I was greatly influenced by his work on Kaluza-Klein supergravity \cite{Freund:1980xh}, like many others. In fact, as put very well in \cite{Duff:1986hr}, in the beginning of the 80's, Peter's paper on Freund-Rubin compactifications of the eleven dimensional supergravity, and another two papers, one by Witten\cite{Witten:1981me} and another by Salam and Strathdee\cite{Salam:1981xd}, started a renaissance in Kaluza-Klein theories. 
Nowadays we take for granted the idea spontaneous compactification, but in the early days of Kaluza-Klein supergravity, moving from dimensional reduction to spontaneous compactification by addressing the underlying dynamics, as highlighted in the title of the Freund-Rubin paper as ``the dynamics of dimensional reduction", and the fact that it works so naturally to give four dimensional spacetime, was a very impactful development. The Freund-Rubin paper not only emphasized this point but it also elevated greatly the stakes for the eleven dimensional supergravity, which has proven to be so important for what has become M-theory since the mid 90's.

In this note, it is fitting to revisit the maximally symmetric Freund-Rubin compactifications of 11D supergravity, namely $AdS_{4/7}\times S^{7/4}$ with $4$-form flux turned on. Firstly, we would like to find out if the resulting Kaluza-Klein spectrum of states can be described in a unified manner. Second, we aim at probing the question of whether singletons and doubleton field equations can be identified in the bulk. 
We will see that a unified treatment of the KK spectra is indeed possible, by exploiting the fact that both of the maximally supersymmetric Freud-Rubin vacua admit analytic continuation to $S^4 \times S^7$. 
As a result, we will see that the detailed supermultiplet structure of the spectrum, as well as the 11D origin of the fluctuations emerges from a simple rule. So far, these spectra have been obtained by separate computations\cite{Englert:1983rn,es1,es2,Casher:1984ym,vanNieuwenhuizen:1984iz,Gunaydin:1984wc}.

It has been observed for the $S^7$ compactification in \cite{es1,es2}, and $S^4$ compactification in \cite{Gunaydin:1984wc}, that the group theoretical structure of the KK spectrum suggests the presence of singletons and doubletons. In \cite{Casher:1984ym}, it was argued that these states vanish identically, while in \cite{es1,es2} they appeared as nonpropagating modes, as the saturated propagator has vanishing residue for the associated poles. 
Examining the issue of whether they can arise as boundary states, we find the linearized field equations for the singletons in $AdS_4$ and doubletons in $AdS_7$, but we also find that there are certain Stuckelberg shift symmetries inherited from the Kaluza-Klein reduction which can be used to gauge them away. Section 5 is devoted to these issues, which are further discussed in the conclusions.   

\section{Preliminaries}

The Freund-Rubin compactifying solutions of 11D supergravity on $AdS_{4/7}\times S^{7/4}$ can be given in a unified fashion such that the only nonvanishing fields are
\bea
{\bar R}_{\mu\nu\rho\sigma} &=& -4 \e \, m^2 ( {\bar g}_{\mu\rho} {\bar g}_{\nu\sigma} - {\bar g}_{\mu\sigma} {\bar g}_{\nu\rho} ) \ ,
\nn\w2
{\bar R}_{\a\b\c\d} &=& \e \, m^2  ( {\bar g}_{\a\c}  {\bar g}_{\b\d} - {\bar g}_{\a\d} {\bar g}_{\b\c} )\ ,
\nn\w2
{\bar F}_{\mu\nu\rho\sigma} &=& 3m\, \ve_{\mu\nu\rho\sigma} \ ,
\label{vac}
\eea
where $m$ is an arbitrary constant, the Levi-Civita tensor $\ve_{\mu\nu\rho\sigma}$ is evaluated in the background, and the following index notation is used
\bea
 \e=+1\, :  \qquad  AdS_4 \times S^7\ , \qquad \mu=0,1,2,3\ ,\qquad \a=1,2,...,7\ , 
 \nn\w2
 \e=-1\, :  \qquad  AdS_7 \times S^4\ , \qquad \a=0,1,...,6\ ,\qquad \mu=1,2,3,4\ , 
 \eea
and accordingly,  $\ve_{\mu\nu\rho\sigma}\,\ve^{\mu\nu\rho\sigma} = -4!\, \e$.
We parametrize the linearized fluctuations around a background as 
\be
g_{MN} = \bar{g}_{MN}+h_{MN}\ , \qquad  A_{MNP} = { \bar A}_{MNP} + a_{MNP}\ .
\ee
Introducing the notation $\Phi = (h_{AB}, a_{CAB})$,  where $A$ is the 11D tangent space index, and coupling to the source ${\cal J}=(T_{AB}, J_{CAB})$, the action quadratic in fluctuations can be written as 
\be
I^{(2)} = \int d^{11}x\, \left(-\frac12 \Phi \cO \Phi  + {\cal J}\Phi\right) \ ,
\ee
where $\cO$ is the wave operator. Since the action is invariant under the background gauge transformations
\be
\d h_{MN} = {\overline\nabla}_{M}\x_{N}+{\overline\nabla}_{N}\x_{M}\ ,\qquad 
\d a_{MNP} = \xi^L {\bar F}_{LMNP} + 3 {\overline\nabla}_{[M} \Lambda_{NP]}\ .
\ee
it follows that the sources must satisfy the constraints (using the normalizations chosen in \cite{es2})
\be
{\overline\nabla}^M T_{MN}+\frac23 {\bar F}_{NPQR} J^{PQR}=0\ ,\qquad {\overline\nabla}^M J_{MNP}=0 \ .
\ee
The following gauge was chosen in \cite{es2} (slightly different from the gauge chosen in \cite{es1})
\be
{\bar \nabla}^M \left( h_{MN} -\frac12 {\bar g}_{MN}\, {\bar g}^{RS} h_{RS} \right) = 0\ ,
\qquad 
{\bar \nabla}^P\,a_{PMN} = 0 \ .
\label{bg1}
\ee
Using the gauge condition, the wave operator $\cO$ can be inverted. Substituting the result into $I^{(2)}$, and importantly using the source constraints, we obtain the {\it saturated propagator}
\be
I^{(2)} = \frac12 \int d^{11}x\, {\cal J}\,\cO^{-1}\, {\cal J}\ .
\ee
This procedure, in the case of ${\rm Minkowski}_4\times S^2$ compactification of $6D$ Maxwell-Einstein theory was employed in \cite{RandjbarDaemi:1982hi}. In that case, the harmonic expansion on the Minkowski spacetime is the usual Fourier transform, while here, where we are dealing with $AdS$ spacetimes, harmonic expansions reduce to those on spheres. After harmonic expansions in the total Euclideanized spacetime, the physical states  are determined from the analysis of the poles in the principle lowest weight, that is the lowest $AdS$ energy $E_0$ plane, in the expression for the saturated propagator. The nonvanishing and positive residues describe the physical states. The manner in which the representation function on spheres and $AdS$ space are related under the analytic continuation was examined in detail in \cite{es1}. It was also shown that the eigenvalues of the second order Casimir operators for the isometry group of the  sphere and $AdS$ space are related by the simple rule where one identifies the leading lowest label with an opposite sign. This simple rule facilitates the physical interpretation of the poles in the lowest weight plane \cite{es1,es2}.

Turning to the Freund-Rubin compactifications of 11D supergravity, both of the maximally supersymmetric vacua can be treated simultaneously by analytically continuing the equations to $S^4 \times S^7$. In this ``democratic '' approach, starting from the universal result on the product of the spheres, one can analytically continue either $S^4$ to $AdS_4$, or with equal ease $S^7$ to $AdS_7$, thereby obtaining not only the group theoretical content of the full spectrum of physical states but also the full information about how they are formed out of the 11D supergravity fields.

\section{Analytic continuation and harmonic expansions on $S_4\times S^7$}

Analytic continuation from $AdS_4\times S^7$ to $S^4 \times S^7$ was described in\cite{es1,es2}. Here we shall formulate it in a way that enables us to analytically continue also $AdS_7\times S^4$  to $S^4 \times S^7$  such that all the harmonic analysis performed on $S^4 \times S^7$ in \cite{es1,es2} are exactly the same as before. 
With this strategy in mind, we consider the metric for $AdS_{d+2}$
\be
ds^2=m^2 ( -\cosh^2\rho dt^2+d\rho^2+\sinh^2\rho\, d\Omega_d) \ ,
\ee
where $d\Omega_d$ is the metric on unit radius $d$-sphere. The Euclideanization rule appropriate for our purposes here is to send %
\be
\rho\mapsto i\rho \ , 
\ee
which gives
\be
ds^2\mapsto - m^2 (\cos^2\rho dt^2+d\rho^2+\sin^2\rho\, d\Omega_d)=-ds_{(E)}^2\ ,
\ee
which is locally the metric of the $(d+2)$-sphere with negative-definite signature. Therefore, in evaluating the linearized field equations around the Euclideanized vacuum solution, we need to send ${\bar g}^{AdS} \to - {\bar g}^{\rm sphere}$ and ${\rm \bar Riem}^{AdS} \to - {\rm \bar Riem}^{\rm sphere}$. In this way, the analytical continuation that enables us to treat both cases simultaneously takes the form
\bea
&& \bar g_{\mu\nu} \to -\e\, \bar g_{\mu\nu}^{S^4}\ ,\qquad \bar g_{\a\b} \to \e\,\bar g_{\a\b}^{S^7}\ .
\eea
In the linearized field equations, there will be $\e$ dependence coming from the Freund-Rubin solution. However, with the analytic continuation prescription described above, the $\e$ factors work out in such a way that the linearized field equations take the same form as those which have already been analyzed for the $AdS_4\times S^7$ vacuum, with the harmonic expansions on $S^4 \times S^7$ fully performed. The treatment of the Levi-Civita symbols requires some care but the key point is that one can take over the results of \cite{es1,es2} and either continue them back to $AdS_4 \times S^7$ as was done in \cite{es2}, or continue back to $AdS_7 \times S^4$ readily. 

Next, we proceed with the harmonic expansions on $S^4 \times S^7 = G/H$ with $G= SO(5)\times SO(8)$ and $H=SO(4)\times SO(7)$. According to the framework described in \cite{Salam:1981xd}, and applied to the case at hand in\cite{es2}, one expands the fluctuations with a given $H$-content in terms of all $G$-representation functions that contain the $H$-representation. The highest weight labelling of the representations is more convenient than the Dynkin labels for this purpose. Using the notation of \cite{es2}, let the highest weight of an H-representation be
\be
H:\qquad (a_1 \,a_2)(b_1\,b_2\,b_3) ;\quad a_1 \ge a_2\ ,\quad b_1\ge b_2\ge b_3\ ,
\label{H}
\ee
All $G$-representations that contain this $H$-representations have the with highest weight
\be
G:\qquad (n\,n_1)(\ell\,\ell_1\,\ell_2\,\ell_3) ;\quad n \ge n_1\ ,\quad \ell \ge \ell_1\ge \ell_2\ge |\ell_3|\ ,
\label{G}
\ee
subject to the conditions 
\be
n \ge a_1 \ge n_1 \ge |a_2|\ ,
\qquad 
\ell \ge b_1 \ge \ell_1 \ge b_2 \ge \ell_2 \ge b_3 \ge  |\ell_3|\ .
\ee
Using these embedding conditions a field $\phi(x,y)$ with a fixed $SO(4)\times SO(7)$ representation $(a_1,a_2)(b_1,b_2,b_3)$ can be expanded in terms of the representation functions of $SO(5)\times SO(8)$ as follows:
\be
\phi_{(a_1a_2)(b_1b_2b_3)}(x,y) = {\rm vol.} (S^4\times S^7) \sum \sqrt{\frac{d_{nn_1} d_{\ell\ell_1\ell_2\ell_3}}{d_{a_1a_2}d_{b_1b_2b_3}}}\,D_{(a_1a_2),p}^{(nn_1)}(L_x^{-1})\,
D_{(b_1b_2b_3),q}^{(\ell\ell_1\ell_2\ell_3)}(L_y^{-1})\,\phi_{pq}^{(nn_2)(\ell\ell_1\ell_2\ell_3)},
\label{me}
\ee
where the summation ranges are as given in \eq{H} and \eq{G}, the $d$'s are the dimensions of the relevant representations, $L_x$ and $L_y$ are the coset representative elements, $D^{(nn_1)}_{(a_1a_2),p}(L_x^{-1})$ is the $(nn_1)$ representation of $L_x^{-1}$, with rows labelled by $(a_1a_2)$ and columns by $p=1,2,..., d_{nn_1}$. The representation matrices(functions) of $L_y^{-1}$ are to be interpreted similarly, and $\phi_{pq}^{(nn_1)(\ell\ell_1\ell_2\ell_3)}$ are $x$ and $y$ independent expansion coefficients. In the computation of saturated propagator, the orthogonality relations for the harmonics are needed. For example, on $S^4$ they take the form
\be
\int_{S^4} d^4x\, \left(\det\,\bar g_{\mu\nu}\right)^{1/2} \, D_{(a),p}^{(n)}\left(L_x^{-1}\right)\,D_{(a),p'}^{(n')}\left(L_x^{-1}\right) = {\rm Vol.}(S^4)\,\frac{d_{(a)}}{d_{(n)}}\,\delta_{pp'} \delta^{(n)(n')}\ ,
\ee
where $(a)$ denotes the row label for the $H$-representations, e.g. $\mu,  (\mu\nu), {\rm etc}$, and $(n)$ is shorthand for $(nn_1)$. Summation over $(a)$ is understood. Similar formula holds on $S^7$.

In these computations repeated use of the following relations are  made
\bea
{\overline\Box} L_x^{-1} &=&  -4m^2 \Big(C_2[SO(5)] -C_2[SO(4)]\Big) L_x^{-1}\ ,
\w2
{\overline \nabla}_\mu \,D_{(a_1a_2),p}^{(nn_2)} (L_x^{-1}) &=& -2m <p|Q_\mu|a_1a_2>\,D_{(a_1a_2),p}^{(nn_2)} (L_x^{-1})\ , 
\label{lemmas}
\eea
where $Q_\mu = M_{\mu 5},\ \mu=1,...,4$ are the $SO(5)/SO(4)$ coset generators. Similar formula hold for the $SO(8)/SO(7)$ coset. The matrix elements $<p|Q_\mu|a_1a_2>$, indeed all matrix elements of $SO(N)$ for any $N$, can be found in \cite{Ottoson'68,Barut:1986dd}, where they are given in Gelfand-Zeitlin (GZ) basis. 
Thus, one needs to find the relation between these basis elements and the tensorial one. Many of these relations can be found in \cite{es2}. 

If one is only interested in determining the KK spectrum of the theory, it is worth noting that there are shortcuts for doing so. In that context, the kind of data provided in Table 1 is very powerful.  Indeed, following the approach of \cite{deBoer:1998kjm}, in the $AdS_4\times S^7$ compactification here, one can start with $(n2)(\ell 000)$ state that describes the graviton tower, and simply compute the tensor product with the supercharge representation $(\1h,\1h)(\1h,\1h,\1h,\pm\1h)$, repeatedly. Comparing with the available representations listed in Table 1, one can deduce the content of Fig 1. The representations that are left over from Table 1 can then be interpreted as being non-propagating. This method works well especially if there is high degree of supersymmetry, and one ``pryamid" of states. For less amount of supersymmetry, one would have to determine the top member of more than one pyramid of states\cite{Deger:1998nm,deBoer:1998kjm}, and repeat the procedure until all supermultiplets are accounted for. However, the details of exactly how the $11D$ fluctuations organize themselves to produce the physical states, and the analysis of possible boundary states may not be available in this approach. 

\section{The spectrum on $AdS_4 \times S^7$ and $AdS_7 \times S^4$ }

With full harmonic expansions on $S^4\times S^7$, the problem of finding the saturated propagator reduces to an algebraic one. The following notation is introduced for the  fluctuations \cite{es1,es2}
\bea
h_{\mu\nu} &=& H_{\mu\nu} + {\bar g}_{\mu\nu}\,M\ ,\qquad {\bar g}^{\mu\nu} H_{\mu\nu} =0\ ,\qquad h_{\mu\a} = K_{\mu\a}\ ,
\nn\w2
h_{\a\b} &=& L_{\a\b} + {\bar g}_{\a\b}\,N\ ,\qquad {\bar g}^{\a\b} L_{\a\b} =0\ , 
\nn\w2
a_{\mu\nu\rho} &=& \ve_{\mu\nu\rho\sigma} W^\sigma\ ,\qquad 
a_{\mu\nu\a}^\pm = \frac12 ( a_{\mu\nu\a}  \pm \frac{i}{2}\, \ve_{\mu\nu}{}^{\rho\sigma}\, a_{\rho\sigma \a} ) \equiv X_{\mu\nu\a}^\pm\ ,
\nn\w2
a_{\mu\a\b} &=& Y_{\mu\a\b}\ ,\qquad a_{\a\b\c} = Z_{\a\b\c}\ .
\label{sdefs}
\eea
The full harmonic expansions on $S^4 \times S^7$ for the bosonic sector give the result \cite{es2}\footnote{We are being cavalier about the overall signs in the individual terms here, with the understand that the sign of the residues at the poles, whether in the $n$-plane, or the $\ell$-plane are always positive, upon properly taking into accounts the rules of the analytic continuations involved.}
\bea
I^{(2)}&=& \sum_{n,\ell}  \Bigg\{ \frac{|T_H^{(n2)(\ell000)}|^2 + |T_L^{(n0)(\ell 200)}|^2 + 4 |J_Y^{(n1)(\ell 110)}|^2}{8 \left(2n+\ell+6\right)\left(2n-\ell\right)} 
\nn\w2
&& + \frac{2 |J_Z^{(n0)(\ell 111)}|^2}{(2n+\ell+9)(2n-\ell-3)} + \frac{2 |J_Z^{(n0)(\ell 11-1)}|^2}{(2n+\ell+3)(2n-\ell+3)} \nn\w2
 &&  + \frac{1}{2\left(2n+\ell+3 \right)\left(2n-\ell-3\right)} \left| \frac{\sqrt{n+1}\,T_K^{(n1)(\ell100)} +4\sqrt{n+2}\,J_X^{(n1)(\ell100)}}{\sqrt{2n+3}}\right|^2 
\nn\\
&& + \frac{8}{\left(2n+\ell+9 \right)\left(2n-\ell+3\right)} \left| \frac{\sqrt{n+2}\,T_K^{(n1)(\ell100)} -4\sqrt{n+1}\,J_X^{(n1)(\ell100)}}{\sqrt{2n+3}}\right|^2
\nn\w2
&&  + I^{(2)}_{\rm scalars} + I^{(2)}_{\rm nonpropagating}  \Bigg\} \ .
\label{ma1}
\eea
The last two terms will be discussed further below. Using this formula, we can continue to either $AdS_4\times S^7$ or $AdS_7 \times S^4$. In the first case, we set $n=-E_0$, and examine the pole in the $E_0$-plane. In the second case, we set $\ell=-E_0$, and look for the poles in the $E_0$-plane, with $E_0$ now denoting the lowest energy in $AdS_7$.

In the first three terms above, one of the poles in the $n$-plane gives $E_0= 3+\frac{\ell}{2}$. They describe towers of physical states with the following $SO(3,2)\times SO(8)$  representation content 
\be
AdS_4 \times S^7\ :\quad 
\begin{cases} H_{\mu\nu}:\quad    (E_0, 2)(\ell 000)\ ,\qquad  E_0=\frac{\ell}{2} +3\ , \quad \ell\ge 0 
\\
Y_{\mu\a\b}: \quad (E_0, 1)(\ell 110)\ ,\qquad  E_0=\frac{\ell}{2} +3\ , \quad \ell\ge 1 
\\ 
L_{\a\b}:\quad\ (E_0,0)(\ell 200)\ ,\qquad  E_0=\frac{\ell}{2}+3\ , \quad \ell\ge 2\ .
\end{cases}
\ee
The second poles are related to the pole discussed above by the replacement $E_0 \to (3-E_0)$, which describe the conjugate representations. We see that $H_{\mu\nu}$ contains the irrep $(3,2)(0000)$ which is the massless graviton, as its energy $E_0=3$ saturates the unitarity bound $E_0=s+1$ for $s=2$. 

From the same saturated propagator \eq{ma1}, it also easy to read of the spectrum in $AdS_7 \times S^4$. To do so, we simply look for the poles in the $\ell$-plane. Those are at $\ell=2n$ and $\ell=-2n-6$, again related to each other by the rule $\ell \to -6-\ell$. Next, we identify $\ell=-E_0$, where $E_0$ now represents the lowest energy in $AdS_7$. Thus, the tower of physical states in this sector are given by
\be
AdS_7 \times S^4\ :\qquad 
\begin{cases} 
L_{\a\b}:\quad \ (E_0, 2,0,0)(n0)\ ,\qquad  E_0=2n+6\ , \quad n\ge 0
\\
Y_{\mu\a\b}: \quad  (E_0, 1,1,0)(n1)\ ,\qquad  E_0=2n+6\ , \quad n\ge 1 
\\ 
H_{\mu\nu}: \quad (E_0, 0,0,0)(n2)\ ,\qquad  E_0=2n+6\ , \quad n\ge 2\ .
\end{cases}
\ee
Now it is the field $L_{\a\b}$ at the bottom floor of the tower with $n=0$ that describes the massless graviton in $AdS_7$ as it has the lowest energy $E_0=6$ that saturates the unitarity bound for the unitary discrete representation of  $SO(6,2)$, as $E_0=\ell_1+2$ with $\ell_1=2$, while $H_{\mu\nu}$ describes a tower of massive scalars. 

Turning to the $(n)(\ell,1,1,\pm 1)$ and $(n0)(\ell100)$ representations, in order to clarify how the poles in the $n$-plane fit into $OSp(8|4)$ multiplets, we relabel  $\ell\to \ell-1$ for the first terms, and $\ell\to \ell+1$ in the second terms in the saturated propagator in this sector. Thus one finds the following towers of physical states
\be
AdS_4 \times S^7\ :\quad 
\begin{cases} Z_{\a\b\c}:\quad\quad\quad\quad\ (E_0^\pm, 0)(\ell\pm 1,0,0,0)\ ,\qquad  E_0^\pm=\frac{\ell}{2} +3 \mp 1\ ,
\\
\left(K_{\mu\a}, X_{\mu\nu\a}^\pm\right) : \quad (E_0^\pm, 1)(\ell\pm 1,1,0,0)\ ,\qquad  E_0^\pm=\frac{\ell}{2} +3 \mp 1\ ,
\end{cases}
\ee
where $\ell\ge 0$ for the upper sign tower, with $\ell=0$ states being the massless scalars in the $35_v$-plet  and massless vectors in the $28$-plet  of $SO(8)$.

In a similar fashion, analytically continuing to $AdS_7$ instead, this time letting $n \to n+1$ for the first terms, and $n\to n-1$ in the second terms discussed above, we easily obtain the following spectrum of states 
\be
AdS_7 \times S^4\ :\quad 
\begin{cases} Z_{\a\b\c}:\quad\quad\quad\quad\ (E_0^\pm, 1,1,\pm 1)(n\pm 1,0)\ ,\qquad  E_0^\pm=2n+6 \mp 1\ ,
\\
\left(K_{\mu\a}, X_{\mu\nu\a}^\pm\right) : \quad (E_0^\pm, 1,0,0)(n\pm 1,1)\ ,\qquad\ \  E_0^\pm=2n+6 \mp 1\ ,
\end{cases}
\ee
where $n\ge 0$ for the upper sign tower, with $n=0$ states being the massless 3-form fields in the $5$-plet, and massless vectors in the $10$-plet  of $SO(5)$, while $n\ge 2$ for the lower sign towers consisting of massive 3-form fields and vectors only. In the first tower, the case of $n=-1$ is special. It will be analyzed in more detail in the next section, where we will see that it describes a doubleton. 

The case of massless 3-form fields also deserves a further comment. In this case, the harmonically expended field equation becomes $(\ell+5)(\ell-5)Z^{(\ell,1,1,-1)(00)}=0$. As shown in \cite{Townsend:1983xs,Pilch:1984xy}, this means that the field equation for this mode factorizes as
\be   
\left(\delta_{\a\b\c}^{\a'\b'\c'}+\frac{1}{12} \ve_{\a\b\c}{}^{\a'\b'\c'\d'}\on_{\d'} \right) \left( \delta_{\a'\b'\c'}^{\a''\b''\c''}-\frac{1}{48} \ve_{\a'\b'\c'}{}^{\a''\b''\c''\d''}\on_{\d''} \right)\,Z_{\a''\b''\c''}^I (y)=0\ ,
\label{PvN}
\ee
where $I=1,...,5$ is the $SO(5)$ vector index. This can be checked by expanding $Z^I_{\a\b\c}(y) = \sum Z_p^{I,(\ell 1 1-1)} D_{\a\b\c,p} (L_y^{-1})$ (we ignore the normalization factors here), and using the relation
\be
\nabla_{[\delta} D_{\a\b\c],p}^{(\ell11-1)} = -\frac{1}{24} (\ell+3)\ \ve_{\a\b\c\delta}{}^{\a'\b'\c'} D_{\a'\b'\c',p}^{(\ell11-1)}\ .
\ee
Recalling the analytical continuation by which $E_0=-\ell$, we see that the first factor in \eq{PvN} gives the lowest energy $E_0=5$ appropriate for the massless $3$-form field, as $E_0=\ell_1+4$ with $\ell_1=1$.

Next we turn to the scalar fields arising in the sector where the fields carry the $(n0)(\ell000)$ representation. This is the most complicated sector as the linearized equatiosn mix the fields $(M,N,\partial W, \partial\partial K, \partial\partial H, \partial\partial L)$. The last two can be eliminated in terms of the remaining ones by means of the gauge conditions. The resulting four coupled linearized equations were analyzed in \cite{es2} where it is found that the residues at two of the resulting poles in the $n$-plane in the saturated propagator vanish, while the other two poles give physical states. To see how these states fit into the supermultiplets, in this case the shifts $\ell\to \ell \pm 2$ are appropriate, and we find the towers
\be
AdS_4 \times S^7\ :\quad (M,N,\partial W, \partial\partial K):\qquad  (E_0^\pm, 0)(\ell\pm 2,0,0,0)\ ,\qquad  E_0^\pm =\frac{\ell}{2}+3\mp 2\ . 
\label{st12b}
\ee
For $\ell=0$ the first tower gives the  massless scalars in the $35$-plet of $SO(8)$. At $\ell=-1$, scalars in $8_v$ of $SO(8)$ reside and have been shown to be gauge modes in \cite{es2}. The question of whether they can be part of a supersingleton boundary supermultiplet will be discussed in the next section.

Analytical continuation from $S^7$ to $AdS_7$ instead, we find that two of the poles give vanishing residue, and the remaining two, upon letting  $n\to n+2$ for one of the poles, and $n\to n-2$ for the other, again for the supermultiplet interpretation, give the towers
\be
AdS_7 \times S^4\ :\quad (M,N,\partial W, \partial\partial K): \qquad (E_0^\pm,0,0,0)(n\pm 2,0)\ ,\qquad  E_0^\pm =2n+6 \mp 2 \ , 
\label{st12c}
\ee
with the upper sign tower starting at $n=0$, and the lower one at $n=2$. The first tower at $n=0$ contains massless scalars $14$-plet of $SO(5)$. At $n=-1$, the there are scalars in $5$-plet of $SO(5)$ which turn out to describe part of the superdoubleton, as we shall see in the next section. 

Finally, we turn to the term $I^{(2)}_{\rm unphysical}$ in \eq{ma1}. This term refers to the remaining sectors:
\be
{\rm Nonpropagating:} \qquad  (n0)(\ell 100)\ , \qquad (n0)(\ell 110)\ ,\qquad (n1)(\ell 000) \ . 
\ee
In the case of analytic continuation to $AdS_4\times AS^7$, it was shown in \cite{es2} that their contribution to the saturated propagator in which the source squared term is divided by a quadratic expression in $\ell$, without any $n$ dependence. Thus, not having a pole in the $n$-plane, these are interpreted as being nonpropagating. In the case of analytic continuation to $AdS_4\times AS^7$, we can easily show that their contribution to the saturated propagator this time has the form of source squared term divided by a quadratic expression in $n$, without any $\ell$ dependence. Thus, not having a pole in the $\ell$-plane, we again see that these states are nonpropagating.

So far we have discussed the bosonic sector of the 11D supergravity. In the fermionic sector, the analytic continuation from $AdS_4\times S^7$ to $S^4 \times S^7$ was presented in \cite{es2} where the complete spectrum, bosonic and fermionic, was worked out. The analytic continuation from $AdS_7 \times S^4$ along the lines described above can be extended to fermionic sector as well, just as in the case of $AdS-4\times S^7$ described in detail in \cite{es2}. 

In the fermionic sector the local supersymmetry transformations of the fluctuations is given by
\be
\delta \psi^A = \on_A \e -\frac{1}{144} \left(\Gamma_A{}^{B_1...B_4} -8\Gamma^{B_1...B_3} \delta_A^{B_4} \right) {\bar F}_{B_1...B_4}\e\ .
\ee
Introducing the source term with a suitable normalization, one finds that local supersymmetry imposes the constraint \cite{es2}
\be
{\overline\nabla}_A J^A -\frac{1}{144} \left(\Gamma_A{}^{B_1...B_4} +8 \Gamma^{B_1...B_3} \delta_A^{B_4} \right) {\bar F}_{B_1...B_4}\,\ve =0\ .
\ee
In \cite{es2} the following gauge is chosen
\be
\Gamma^A \psi_A=0\ .
\label{fg1}
\ee
Writing $\Gamma^r =\gamma^r \times 1$, and $\Gamma^i=\gamma_5 \times \gamma^i$, where the 11D tangent space index is split as $A=(r,i)$, with $r=0,1,2,3$ and $i=4,..,10$, and defining the fluctuations fields 
\be
\psi_r=\left(\frac{1+i\gamma_5}{\sqrt 2}\right) \left(\eta_r+\gamma_r \lambda\right)\ ,\qquad 
\psi_i=\left(\frac{1+i\gamma_5}{\sqrt 2}\right) \left(\chi_i+\gamma_i \theta\right)\ ,
\ee
where $\eta_r$ and $\chi_i$ are $\gamma$-traceless, and the $\gamma_5$ dependent prefactors are introduced for convenience.
We shall skip the details of the analytic continuation in this sector, as it is similar to the one described in \cite{es2}. The procedure is exactly as the one explained for the bosonic sector, and the saturated propagator for this sector provided in \cite{es2} yields the result
\be
AdS_4 \times S^7\ :\quad 
\begin{cases} \eta_r :\quad \quad\quad\quad \quad  \left(E_0^\pm, \3h\right)\left(\ell\pm\1h,\1h,\1h,\pm\1h\right)\ ,\qquad  E_0^\pm =\frac{\ell}{2} + 3 \mp \1h\ , 
\\
\chi_i : \quad\quad\quad\quad\quad  \left(E_0^\pm, \1h\right)\left(\ell\pm\1h,\3h,\1h,\mp\1h\right)\ ,\qquad  E_0^\pm =\frac{\ell}{2} +3 \mp \1h\ , 
\\ 
\left(\partial \eta, \partial\chi, \lambda\right) :\quad\ \left(E_0^\pm, \1h\right)\left(\ell\pm\3h,\1h,\1h,\mp\1h\right)\ ,\qquad  E_0^\pm =\frac{\ell}{2} +3 \mp \3h\ .
\end{cases}
\ee
In the first equation, the $\ell=0$ states are the massless gravitini in the $8_s$ of $SO(8)$. In the last result above, three coupled linearize equations were analyzed in \cite{es2}, and additional poles in the saturated propagator were shown to give vanishing residue; hence only the towers displayed above arise as physical.  Note also the shifts of $\ell$ by $\pm\1h$ or $\pm\3h$, again for the purposes of supermultiplet interpretation; see Figure 1. Furthermore, $\ell=-1$ in the last tower gives fermions which were shown to be gauge modes in \cite{es2}. Whether they can survive as the fermionic partner of a supersingleton will be examined in the next section. 

Using the saturated propagator given in \cite{es2} this time to continue analytically to $AdS_7\times S^4$ instead, we easily obtain 
the result
\be
AdS_7 \times S^4\ :\quad 
\begin{cases} \chi_i :\quad \quad\quad\quad \quad  \left(E_0\pm \1h,\3h,\1h,\mp \1h\right)\left(n\pm\1h,\1h \right)\ ,\qquad  E_0 =2n+6 \mp \1h\ ,
\\
\eta_r : \quad\quad\quad\quad\quad \left(E_0^\pm,\1h,\1h,\pm\1h\right)\left(n\pm\1h,\3h \right)\ ,\qquad\quad\ \  E_0^\pm =2n+6\mp\1h\ , 
\\ 
\left(\partial \eta, \partial\chi, \lambda\right) :\quad\ \left(E_0^\pm,\1h,\1h,\mp\1h\right)\left(n\pm \3h,\1h\right)\ ,\qquad\quad\ \ \ E_0^\pm =2n+6 \mp \3h\ .
\end{cases}
\ee
In the first equation, the upper sign tower starts at $n=0$, which is the {\it massless} gravitino. In the last equation $n=-1$ is an acceptable representation, and it will be shown in the next section to have the appropriate field equation for a fermionic partner of a superdoubleton.

In summary, the results for the full spectrum in the $AdS_4 \times S^7$ and $AdS_7\times S^4$ are given in Fig. 1 and Fig. 2. In \cite{es2}, it was observed that the following relation holds for a particular combination of the second order Casimir operator eigenvalues at each level:
\be
2C_2[SO(3,2)]+C_2[SO(8)]= \frac32 \left(\ell+2\right)\left(\ell+4\right)\ .
\label{cr1}
\ee
In a similar fashion, here we find that the following relation holds in the case of $AdS_7 \times S^4$ Kaluza-Klein spectrum organized into levels labeled by $n$:
\be
2C_2[SO(5)]+C_2[SO(6,2)]= 6\left(n+1\right)\left(n+2\right)\ .
\label{cr2}
\ee
Interestingly, this vanishes for the doubleton multiplet for which $n=-1$.

\section{Search for singletons and doubletons}

Fig 1 shows the full spectrum of 11D sugra compactified on $AdS_4 \times S^7$ compactification, in such a way that each value of $ \ell=0,1,2...,$ represents an $OSp(8|4)$ multiplet. For $\ell=0$ one has the massless maximal 4D supergravity multiplet, and the rest are massive supermultiplets. In \cite{es1,es2}, while it was shown that the representations for $\ell=-1$ are gauge modes, it was observed that these form the singleton supermultiplet of $OSp(8|4)$\footnote{The conjecture for its existence in the spectrum, appeared in the first reference in \cite{es1}.}. The question arises as to whether the gauge fixing procedure allows their existence as boundary states. We begin by noting that for $\ell=-1$ (corresponding to $\ell=1$ before the shift $\ell\to \ell+2$ that defines the universal KK level $\ell$), the scalar $\partial\partial K$ can be eliminated by using the gauge condition, and this leads to field equations for the scalars $(M,N,\partial W)$, which we recall are the fields $h_\mu{}^\mu,\, h_\a{}^\a$ and $\ve^{\mu\nu\rho\sigma} \left(F_{\mu\nu\rho\sigma}\right)^{\rm lin.}$. Using the results given in \cite{es2}, one then finds a linear combination of their equations of motion that takes the form\footnote{For a detailed analysis of the singleton field equations, see \cite{Flato:1980zk}.} (henceforth  all covariant derivatives are understood to be evaluated in the background):
\be
\ell=-1\,: \qquad \left(\Box_{AdS_4} + 5 m^2\right)\left( 6M-12N-\nabla_\mu W^\mu\right)^I =0\ ,
\label{fff1}
\ee
where we have re-introduced the parameter $m=1/(2L_{AdS_4})$, and $I=1,...,8$ labels the $(1,0,0,0)$ representation of $SO(8)$. 
This is the appropriate field equation for a singleton with lowest energy $E_0=\frac12$. 
After examining the fermionic sector, we shall come back to the question of whether the boundary states described by this equations survive the fixing of local symmetries.

In the fermionic sector the candidate singleton carries the representation $(1,\1h)(\1h,\1h,\1h,-\1h)$. To better understand the role of supersymmetry gauge fixing, let us not impose the gauge condition \eq{fg1} to begin with. Thus, we examine the $\gamma$-trace of the $4+7$ split of the linearized gravitino equation prior to any gauge fixing. Making use of the fact that for $\ell=\1h$ we have $\nabla_i\chi^i=0$ \cite{es2}, these equations determine $\nabla_r\eta^r$ in terms of $(\lambda, i\theta)$, and furthermore give
\be
\left(\slashed{\nabla}_4-\frac{12}{7} i \slashed{\nabla}_7 +2\right) i\theta (x,y) -\frac47 \left(i\slashed{\nabla}_7+ \frac72\right)\lambda (x,y)=0\ ,
\label{neom}
\ee
and the linearized supersymmetry transformations take the form
\be
\delta\lambda (x,y) = \ft14 (\slashed{\nabla}_4-4 ) \epsilon (x,y)\ ,\qquad \delta i\theta(x,y) = \ft17 (i\slashed{\nabla}_7 +\ft72) \epsilon (x,y)\ .
\label{sn1}
\ee
Denoting any of the spinors occurring above generically by $\psi(x,y)$, it is understood that its harmonic expansion on $S^7$ is of the form $\psi(x,y)= \psi_+(x) D_+ (L_y^{-1}) + \psi_-(x) D_- (L_y^{-1})$, where $D_\pm(L_y^{-1})$ denote the $SO(8)$ representation functions in $(\1h,\1h,\1h,\pm \1h)$ obeying $ i\slashed{\nabla}_7 D_\pm(L_y^{-1}) = \mp\frac72 D_\pm (L_y^{-1})$. Thus equations \eq{neom} and \eq{sn1} give
\be
(\slashed{\nabla}_4 -4)i\theta_-(x) -4 \lambda_-(x)=0\ , 
\label{le1}
\ee
invariant under\footnote{Correcting the sign of the last term in the variation of the gravitino given in eq. (3) of \cite{es2}.}
\be
\delta i\theta_-(x) = \epsilon_-(x)\ , \qquad \delta\lambda_-(x)=\ft14 \left(\slashed{\nabla}_4 -4\right)\epsilon_-(x)\ ,
\ee
and  $(\slashed{\nabla}_4 +8)\, \theta_+(x)=0$ with $\delta \theta_+(x)=0$. Harmonic expansion of $\theta_+(x)$ on $AdS_4$ gives the lowest energy $E_0=\frac{11}{2}$, thus describing the physical state at level $\ell=2$ shown in Fig 1. As for $\theta_-(x)$, it can be gauged away by using the parameter $\epsilon_-(x)$, in which case $\lambda_-(x)$ vanishes by its field equation. However, suppose we fix teh following gauge instead
\be
4\lambda +i\beta\theta_-(x)=0\ ,
\ee
where $\beta$ is a constant parameter. The, the field equation becomes
\be
(\slashed{\nabla}_4 -4 + \beta)\,\theta_-(x)=0\ .
\ee
This has a solution with $AdS$ lowest energy $E_0=\frac12(\beta-1)$, with the unitarity bound imposing the condition $\beta\ge 3$. Interestingly, the choice $\beta=3$ which saturates the unitarity bound gives the singleton field equation\footnote{The gauge condition \eq{fg1} instead gives $\left(\slashed{\nabla}_4 +3\right) \theta =0$, yielding the lowest energy  $E_0=3$ \cite{es2}. This field equation arises in the $N=1$ supersymmetric Wess-Zumino model in $AdS_4$ \cite{Breitenlohner:1982jf}, where the states $(\frac52,0)+(3,\frac12)+(\frac72,0)$ form a massive scalar multiplet. The free action for this case is given in eq. (B1) of \cite{Breitenlohner:1982jf} with $\mu=\frac32$. In \cite{Nicolai:1984gb}, the value $\mu=\frac32$ was mentioned as accommodating a boundary $N=8$ supersingleton. We correct that statement here by noting that  it should have read $\mu=\frac12$. I thank Yoshiaki Tanii for pointing this out, and also for noting that the  supersingletons for $\mu=\frac12$ is related to the one for $\mu=-\frac12$ by a field redefinition.}. %
On the other hand, maintaining the gauge condition imposes the condition involving the same wave operator, namely, $(\slashed{\nabla}_4 -4 + \beta)\,\epsilon_-(x)=0$. 
Since the field equation satisfied by the residual symmetry parameter $\e_-(x)$ coincides with that of the fermionic field $\theta_-(x)$ for any value of $\beta$, it follows that the latter can be removed  entirely by using this residual symmetry, again for any value of $\beta$. Therefore, even though $\beta=3$ gives the singleton field equation for $\theta_-(x)$ this field can nonetheless be removed entirely by fixing the Stuckelberg symmetry. By supersymmetry, we expect  that similar phenomenon must be present for the bosonic singleton equation \eq{fff1} as well, namely the KK reduction of the $11D$ general coordinate and tensor gauge transformations must provide the required residual Stuckelberg shift symmetries to remove them. The nature of these symmetries is similar to those described in detail in \cite{Deger:1998nm} for $6D$ supergravity on  $AdS_3 \times S^3$. 


Let us now examine the linearized field equations for $n=-1$ (corresponding to $n=\1h$ after the relabelling $n \to n+\3h$ to define a universal KK level number $n$) in the $AdS_7\times S^4$ compactification. Again one can eliminate $\partial\partial K$, and using the results of \cite{es2} one finds that a particular linear combination of the  field equations for $(M,N,\partial W)$ in $AdS_7$ takes the form 
\be
n=-1\,: \qquad \left(\ob_{AdS_7} +8 m^2\right) \left(5M+28N-6\nabla_\mu W^\mu\right)^I=0\ ,
\label{deom}
\ee
where $m=1/L_{AdS_7}$ and $I=1,...,5$ labels the vector representation of $SO(5)$. This equation admits a solution with lowest energy $E_0=2$ which is appropriate for a doubleton scalar. The only other bosonic state at $n=-1$ in Fig 2, carries the $SO(6,2)\times SO(5)$ representation $(3,1,1,-1)(0,0)$. Its field equation is that of $Z_{\a\b\c}$ expanded on $S^4$ with the $n=-1$ mode kept. The result is 
\be
n=-1\,:   \qquad \left(\ob_{AdS^7}  +12 m^2 \right) Z_{\a\b\c} +\ve_{\a\b\c}{}^{\a'\b'\c'\d'}\, \on_{\a'}\, Z_{\b'\c'\d'}=0\ ,
\label{deom2}
\ee
where $Z_{\a\b\c}$ depends only on the $7D$ coordinates, and it is a singlet of $SO(5)$. This equation factorizes as
\be   
n=-1\,: \qquad  \left( \delta_{\a\b\c}^{\a'\b'\c'} + \frac{1}{36} \, \ve_{\a\b\c}{}^{\a'\b'\c'\d'}\on_{\d'} \right)\, \left( \ve_{\a'\b'\c'}{}^{\a''\b''\c''\d''}\on_{\d''}\right)\, Z_{\a''\b''\c''} =0\ ,
\label{doubleton2}
\ee
The general solution is a linear combination of those annihilated by the first or second first order wave operator, the second one giving the lowest energy $E_0=3$ state which is appropriate for the doubleton representation. 

Turning  to fermions, at $n=-1$, making use of the fact that $\nabla_r \eta^r=0$ in this sector, the $\gamma$-trace of the linearized gravitino equation of motion on $AdS_7 \times S^4$, prior to any gauge fixing, now determines $\nabla_i \chi^i$ in terms of $(\theta, i\lambda)$, and furthermore gives
\be
\left(\slashed{\nabla}_7+\frac32 i \slashed{\nabla}_4 +\frac52\right) i\lambda(y,x) -\frac74 \left(i\slashed{\nabla}_4 - 4\right)\theta (y,x)=0\ ,
\label{neom2}
\ee
invariant under
\be
\delta \theta (y,x) = \ft17 (\slashed{\nabla}_7 -\ft72) \epsilon (y,x)\ ,\qquad \delta i \lambda (y,x) = \ft14 (i\slashed{\nabla}_4-4 ) \epsilon (y,x)\ .
\label{sn2}
\ee 
Denoting any of the spinors occurring above generically by $\psi(y,x)$, it is understood that its harmonic expansion on $S^4$ is of the form $\psi(y,x)= \psi_+(y) D_+ (L_x^{-1}) + \psi_-(y) D_- (L_x^{-1})$, where $D_\pm(L_x^{-1})$ are the $SO(5)$ representation of the $SO(5)/SO(4)$ coset representative elements $L_x^{-1}$ with the row labeled by $(\1h, \pm \1h)$ representation of $SO(4)\subset SO(5)$, and the column by the $(\1h,\1h)$ representation of $SO(5)$. They obey $ i\slashed{\nabla}_4 D_\pm = \mp 4 D_\mp$. Thus equations \eq{neom2} and \eq{sn2} give
\be
(\slashed{\nabla}_7 -\ft72)i\lambda_-(y) +14\, \theta_-(y)=0\ , 
\label{le2}
\ee
invariant under
\be
\delta i\lambda_-(y) = -2\epsilon_-(y)\ , \qquad \delta\theta_-(y)=\ft17 \left(\slashed{\nabla}_7 -\ft72\right)\epsilon_-(y)\ ,
\ee
and  $(\slashed{\nabla}_7 +\frac{17}{2}) \lambda_+(y)=0$ with $\delta \lambda_+(y)=0$. In the latter equation, harmonic expansion of $\lambda_+(y)$ on $AdS_7$ gives the lowest energy $E_0=\frac{23}{2}$ which is the physical state at  level $n=2$ shown in Fig 2. As for $\lambda_-(y)$, it can be gauged away by using the parameter $\epsilon_-(x)$, in which case $\theta_-(y)$ vanishes by its field equation. However, if we choose the following gauge condition
\be
 7i\theta_-(y) + \tilde\beta \lambda_-(y)=0\ ,
\ee
the fermionic field equation becomes
\be
\left(\slashed{\nabla}_7 -\ft72 + 2\tilde\beta \right)\,\lambda_-(y)=0\ , 
\label{le3}
\ee
which gives the lowest energy $E_0=\frac12(13-4\tilde\beta)$ with the unitarity bound requiring $\tilde\beta\ge 2$. Saturating this bound by taking $\tilde\beta=2$ gives the fermionic doubleton field equation yielding the $AdS_7$ lowest energy $E_0=\frac52$ solution. Maintaining the gauge condition imposes the constraint $\left(\slashed{\nabla}_7 -\ft72 + 2\tilde\beta \right)\,\epsilon_-(y)=0$.Thus, the picture which emerges here is similar to the one we encountered for the singletons in $AdS_4$, and we see that the residual symmetries can be used to remove entirely the field $\lambda_(y)$ for any value of $\tilde\beta$, even though it satisfies the singleton field equation for $\tilde\beta=2$. By supersymmetry, we deduce that the fields which we found to obey the doubleton field equations above can also be removed by residual Stuckelberg shift symmetries coming from the KK reduction of the $11D$ general coordinate and tensor gauge transformations.

\section{Conclusions}

We have found a simple rule that relates the spectrum of physicals stats in the Freund-Rubin compactifications of 11D supergravity on $AdS_4\times S^7$ and $AdS_7\times S^4$. Thus, from the $SO(3,2)\times SO(8)$ lowest weights of the spectrum in the $AdS_4 \times S^7$ compactification given by
\be
AdS_4 \times S^7\,: \qquad (\ve_0\mp  a, n_1)(\ell\pm a, \ell_1, \ell_2, \ell_3)\ ,\qquad \ve_0 :=\frac{\ell}{2}+3\ ,\quad a=0,\ft12, 1, \ft32, 2\ ,
\ee
we deduce the $SO(6,2)\times SO(5)$ lowest weights in the $AdS_7 \times S^4$ compactification by a remarkably simply rule that gives 
\be
AdS_7 \times S^4\,:\qquad  (\e_0\mp a, \ell_1, \ell_2, \ell_3)(n\pm a, n_1)\ ,\qquad \e_0 :=2n+6 \ ,\quad a=0,\ft12, 1, \ft32, 2\ .
\ee
The rule is to interchange  the spacetime and internal symmetry labels such that $(\ve_0, \ell, a)$ goes to $(n, \e_0,  -a)$.
In Fig 1 and Fig 2 each value of the integer $\ell\ge 0$ in Fig 1, and $n\ge 0$ in Fig 2, describe the states which form supermultiplets of $OSp(8|4)$ and $OSp(6,2|4)$, respectively (in the latter case see \cite{DHoker:2000pvz,Beccaria:2014qea} for the AdS lowest energies). The $11D$ origin of the states, and how they transform to each other is also displayed in these figures. 

Possible uses of the simple spectral relation given above, other than providing the complete spectrum for one background from the spectrum for another background, and the relations \eq{cr1} and \eq{cr2}, remain to be seen. For example, the vanishing of Casimir energies that has been shown for $AdS_4\times S^7$ in \cite{Gibbons:1984dg,Inami:1984vp} and for $AdS_7 \times S^4$ in \cite{Beccaria:2014qea}, may possibly be understood from a different angle afforded by these relations. One may also investigate whether the spectral relationship of the kind presented here exists for other compactifications as well\cite{Corbino:2017tfl}.

For the purposes of finding the KK spectrum, it is sufficient to perform the full harmonic expansions and determine the saturated propagator in the space of lowest weights. However, for some other field theoretic purposes, one may need to construct the 2-point function $\Delta(x,x')$ in the coordinate space. To do this, one may employ a Sommerfeld-Watson transformation but we shall not pursue that here. The computation of interactions is also of considerable interest in the context of consistent KK truncation schemes and holography \cite{DHoker:2000pvz}. 

In this paper, we have also found the linearized field equations for the singletons and doubletons in the bulk. However, these turns out to be gauge dependent results, and we have shown that  residual Stuckelberg shift symmetries inherited from the Kaluza-Klein reduction can be used to remove them. 
We have displayed these symmetries explicitly for the fermions but they are expected to arise in a similar fashion in the bosonic sector as well, as described in detail in \cite{Deger:1998nm} in the context of $6D$ supergravity on $AdS_3 \times S^3$. 
These gauge symmetries are not to be confused with the $AdS$ symmetry that operates on the solution space. In the latter case, as explained in detail in \cite{Flato:1986up,Flato:1980zk}, there is a sense in which the singletons can be treated in the framework of a gauge theory in which the solution space, after modding out by gauge transformations that fall off rapidly in the direction of spatial infinity, does support the singletons as boundary states. What we have seen in the Freund-Rubin compactification of $11D$ supergravity is that there is an additional local Stuckelberg symmetry coming from $11D$, other than the $AdS$ symmetry of the background, which removes these states.  

The general expectation that singletons have a role to play in AdS/CFT holography \cite{Aharony:1999ti,Maldacena:2001xj}, motivates a further study of singletons in the context of KK supergravity. The arguments that have been given in support of their presence tend to involve $BF$ type bosonic topological field theories in the bulk. In particular, a detailed study of the $AdS_5$, and a general discussion of the BF type theories in this context exists (see  \cite{Maldacena:2001xj}, and references therein). However, the way  in which suitable topological field theories may arise from the flux compactifications and how coupling to supergravity may occur apparently has not been investigated so far. 
The fact that the BF type theories considered involve p-forms of supergravities, and their behaviour on the boundaries plays an essential role, suggests that the 3-form potential arising in the singleton and doubleton field equations we have found may involve some global considerations that make them survive on the boundary, despite the presence of the Stuckelberg shift symmetries. Whether this is the case remains to be investigated.

\section*{Acknowledgements}

I would like to thank Juan Maldacena, Hermann Nicolai, Henning  Samtleben and Yoshiaki Tanii for useful discussions. I am very grateful to John Strathdee, to whom the papers \cite{es1,es2} owe tremendously. This work is supported in part by NSF grant PHY-1803875. 

\appendix


\ta 




\fa

\phantom{xxx}

\newpage


 

\fb

\newpage


\end{document}